\documentclass[journal]{IEEEtran}
\usepackage{cite}
\usepackage{color,soul}
\usepackage{gensymb}
\usepackage{dsfont}
 \usepackage{bbm}

\ifCLASSINFOpdf
  \usepackage[pdftex]{graphicx}
  \graphicspath{{../pdf/}{../jpeg/}}
  \DeclareGraphicsExtensions{.pdf,.jpeg,.png}
\else
  \usepackage[dvips]{graphicx}
  \graphicspath{{../eps/}}
  \DeclareGraphicsExtensions{.eps}
\fi
\usepackage[cmex10]{amsmath}

\usepackage{amssymb}
{}

\usepackage{euscript}

\usepackage{multirow}
\usepackage{algorithm}
\usepackage{algpseudocode}
\usepackage{diagbox}
\usepackage{physics}
\usepackage{bm}
\usepackage{tikz}
\usepackage{comment}

\usepackage{amsmath,lipsum}
\usepackage{cuted}

\usepackage{lipsum}
\usepackage{multicol}
\usepackage{comment}

\usepackage{xcolor,colortbl}
\definecolor{myred}{RGB}{205 38 38}
 \usepackage{bbm}
\definecolor{myblue}{RGB}{105,89,205}
\definecolor{myorange}{RGB}{238,92,66}
\definecolor{mygray}{RGB}{205,201,201}

\ifCLASSOPTIONcompsoc
  \usepackage[caption=false,font=normalsize,labelfont=sf,textfont=sf]{subfig}
\else
  \usepackage[caption=false,font=footnotesize]{subfig}
\fi
\usepackage{url}
\allowdisplaybreaks[4]

\usepackage{enumitem}

\begin{document}
%
\title{Tractable Data Enriched Distributionally Robust Chance-Constrained CVR}
%
%
%

\author{
         Qianzhi Zhang,~\IEEEmembership{Graduate Student Member,~IEEE,}
         Fankun Bu,~\IEEEmembership{Graduate Student Member,~IEEE,}\\
         Yi Guo,~\IEEEmembership{Member,~IEEE,}
         and Zhaoyu Wang,~\IEEEmembership{Senior Member,~IEEE,}
\thanks{
This work was supported in part by the U.S. Department of Energy Wind Energy Technologies Office under Grant DE-EE0008956, and in part by the National Science Foundation under ECCS 1929975. This work was partially supported by an ETH Z\"{u}rich Postdoctoral Fellowship. (\emph{Corresponding author: Zhaoyu Wang}).

Qianzhi Zhang, Fankun Bu, and Zhaoyu Wang are with the Department of Electrical and Computer Engineering, Iowa State University, Ames, IA 50011 USA (e-mail: qianzhi@iastate.edu; fbu@iastate.edu; wzy@iastate.edu).

Yi Guo is with Power Systems Laboratory at ETH Z\"{u}rich, Z\"{u}rich, 8092, Switzerland (e-mail: guo@eeh.ee.ethz.ch).}}

\maketitle

\begin{abstract}
This paper proposes a tractable distributionally robust chance-constrained conservation voltage reduction (DRCC-CVR) method with enriched data-based ambiguity set in unbalanced three-phase distribution systems. The increasing penetration of distributed renewable energy not only brings clean power but also challenges the voltage regulation and energy-saving performance of CVR by introducing high uncertainties to distribution systems. In most cases, the conventional robust optimization methods for CVR only provide conservative solutions. To better consider the impacts of load and PV generation uncertainties on CVR implementation in distribution systems and provide less conservative solutions, this paper develops a data-based DRCC-CVR model with tractable reformulation and data enrichment method. Even though the uncertainties of load and photovoltaic (PV) can be captured by data, the availability of smart meters (SMs) and micro-phasor measurement units (PMUs) is restricted by cost budget. The limited data access may hinder the performance of the proposed DRCC-CVR. Thus, we further present a data enrichment method to statistically recover the high-resolution load and PV generation data from low-resolution data with Gaussian Process Regression (GPR) and Markov Chain (MC) models, which can be used to construct a data-based moment ambiguity set of uncertainty distributions for the proposed DRCC-CVR. Finally, the nonlinear power flow and voltage dependant load models and DRCC with moment-based ambiguity set are reformulated to be computationally tractable and tested on a real distribution feeder in Midwest U. S. to validate the effectiveness and robustness of the proposed method. 
\end{abstract}


\IEEEpeerreviewmaketitle

\section{Introduction}
\IEEEPARstart{C}{onservation} voltage reduction (CVR) can reduce the voltage for peak load shaving and long-term energy-saving \cite{CVR_2}. To achieve system-wide optimal performance, voltage/var optimization-based CVR (VVO-CVR) is previously studied \cite{CVR_2,CVR_3,CVR_4,CVR_5}, which can be cast into an optimal power flow  program. While the previous works have contributed valuable insights to VVO-CVR, there are problems remaining open, summarized as follows:

(1) \textit{The impact of load and renewable uncertainties on VVO-CVR:} 
In \cite{CVR_3}, a linear least-squares centralized optimization model is developed for coordinating combinations of voltage regulating devices and PVs to implement CVR in distribution systems. In \cite{CVR_4} and \cite{CVR_5}, several alternating direction method of multipliers (ADMM)-based distributed optimization algorithms are developed for CVR implementation, which can decompose a large-scale VVO-CVR problem into several small-scale problems with improved scalability. However, the above centralized or distributed VVO-CVR works \cite{CVR_3,CVR_4,CVR_5} are developed based on deterministic optimization methods, which assume the perfect forecasts of load and renewable power. Neglecting those prediction errors could result in potential violations of operational constraints, such as bus voltage constraints in VVO-CVR. To consider the impacts of load and renewable uncertainties on voltage regulation for CVR implementation, stochastic programming (SP) and robust optimization (RO) are applied in some existing works \cite{SP_VVC_1,SP_VVC_2,RO_VVC_1,RO_VVC_2}. In \cite{SP_VVC_1} and \cite{SP_VVC_2}, the scenario-based SP methods aim at optimizing reactive power dispatch for voltage regulation with expected performance and an accurate probability distribution model. While the SP methods need an accurate probability distribution model and a large sampling number of scenarios, it requires heavy computational efforts. In \cite{RO_VVC_1} and \cite{RO_VVC_2}, the RO methods are developed to handle the uncertain load and renewable power production. However, the RO methods can only give a feasible solution for the worst-case scenario, which is too conservative and hinders the performance of CVR. 

Recently, the distributionally robust optimization (DRO) has been considered a more effective way to handle uncertainty in power systems, such as economic and power dispatch \cite{DRO_2} and \cite{DRO_3}. The DRO can construct an ambiguity set of probability distributions based on historical data, including all possible uncertainty distributions. Thus, the DRO can ensure constraints are satisfied for any distributions in the ambiguity set built upon distribution moments or structure information. Also, the distributionally robust chance-constrained (DRCC) models \cite{DRCC_1} and \cite{DRCC_2} are developed in power systems, which integrate chance constraints to enforce certain events within a probability threshold. Even though DRO and DRCC models have some advantages over conventional deterministic and stochastic approaches, there are still challenges to use the DRCC method for CVR implementation, including how to formulate DRCC in a tractable way for CVR implementation and how to construct an ambiguity set with limited access to historical and real-time load and PV generation data.    

(2) \textit{Intractable DRCC and nonlinear voltage-dependent load models:} 
The VVO-CVR problem is nonlinear and intractable, which makes a distributionally robust stochastic reformulation of the VVO-CVR problem even more challenging. In \cite{DRO_Apx_1}, an approximation method is proposed for conventional DRO problems, while the approximation for DRCC problems is not considered. Also, the loads are considered as voltage-independent models in above reference of DRO and DRCC \cite{DRO_2,DRO_3,DRCC_1,DRCC_2}. In practice, the nature of CVR is lowering network voltages to reduce the voltage, and the literature on the CVR problem validates the necessity of voltage-dependent loads, such as ZIP load and exponential load models \cite{CVR_4}. In existing research works \cite{ZIP_Apx_1} and \cite{ZIP_Apx_2}, some approximation methods are applied to linearize ZIP load model in deterministic optimization. However, the linearization and convexification for ZIP load model in DRO or DRCC with uncertainties of load are not considered in previous works \cite{DRO_Apx_1,DRO_2,DRO_3,DRCC_1,DRCC_2,ZIP_Apx_1,ZIP_Apx_2}. 

(3) \textit{Availability of data for constructing ambiguity set of load and renewable uncertainties:} 
Even though the ambiguity set of uncertainty in DRCC can characterize a group of possibility distributions, defining a high-quality ambiguity set is non-trivial, as one needs to decide the trade-off between the conservativeness of decisions and the operational efficiency, while considering the mathematical tractability. To construct an ambiguity set of uncertainties in DRCC, the conventional way is using statistical inference and data analysis methods with the historical data and system feedback measurements. For example, the existing DRO works \cite{Data_DRO_2} and \cite{Data_DRO_3} use data-based method to construct ambiguity set. However, the aforementioned data-based ambiguity set relies heavily on either sufficiently high-resolution data or complex machine learning methods.

To capture the uncertainties of load and PV generation through data-based methods in modern distribution networks, micro-phasor measurement units (PMUs) and smart meters (SMs) are implemented to record load and PV generation data, where micro-PMUs\footnote{Micro-PMUs are synchrophasor devices that high-speed record real-time stamped data measurement of power and energy consumption. Micro-PMUs have a high sampling rate, e.g., one sample per second or higher \cite{data_enrich}} can record high-resolution data (1-second resolution or higher), and SMs\footnote{SMs are electronic devices that record power and energy consumption and can communicate remotely with utility. SMs have a relatively low sampling rate compared to micro-PMUs, e.g., one sample per hour \cite{data_enrich}.} can record low-resolution data (typically 1-hour resolution). However, due to the cost issue, micro-PMUs are only installed at limited locations in real distribution networks, while SMs are widely installed in real distribution networks. Therefore, the access to available data for constructing an ambiguity set is limited by the number of micro-PMUs and SMs. If we only use limited data to construct an ambiguity set, it may hinder the performance of the DRCC program. The data from SMs and micro-PMUs are necessary but need modification to support our approaches. Therefore, we need to enrich the load and PV generation data, then construct the data-based ambiguity set for uncertainties of load and PV generation.

To address the above challenges, this paper proposes a tractable data-based DRCC-CVR model under the uncertainties of voltage-dependent load and PV generation. Inspired by our previous work \cite{data_enrich}, we apply the data enrichment method to both load and PV generation data from SMs and micro-PMU, then construct the ambiguity set of uncertainties with the enriched data for the proposed DRCC-CVR. This data enrichment method enables a strong connection between the moment-based representation of the limited data set and the moment-based ambiguity set of the DRCC model, which avoids over- or under-conservativeness of the decisions. The main contributions of this paper are three-fold: 
\begin{itemize}
\item \textit{DRCC-CVR model with uncertainties of load and solar PV generation:} To consider the impacts of load and renewable uncertainties on voltage regulation and energy-saving performance of CVR in the unbalanced three-phase distribution systems, we present a deterministic VVO-CVR model and extend it by introducing chance constraints for possible voltage violations due to the load and the renewable uncertainties. To obtain proper conservative solutions robust to the high-resolution dataset and ensure a better performance for energy-saving and voltage regulation of CVR, we further integrate the chance constraints with DRO techniques to propose a DRCC-CVR model.    
\item \textit{Linearization and tractable reformulation of DRCC-CVR model:} To make our proposed DRCC-CVR model tractable, we present the linearized version of the power flow model and voltage-dependent ZIP load model. Then, we reformulate the chance constraints with a moment-based ambiguity set of load and PV generation uncertainties in a tractable way.
\item \textit{Data-enriching moment-based ambiguity set with SM and micro-PMU data:} To guarantee the performance of the data-based DRCC-CVR model, we leverage the data enrichment method to recover high-resolution load and PV generation data from SMs and micro-PMUs. The moment information of load and PV generation uncertainties are execrated from the enriched data to construct the moment-based ambiguity sets for the proposed DRCC-CVR model. 
\end{itemize}


\section{The Problem Formulation and the Proposed Method}\label{sec:framework}
\subsection{Solving a VVO-CVR Problem in the Unbalanced Three-phase Distribution Networks}
In this paper, we consider an unbalanced three-phase radial distribution network that consists of $N$ buses denoted by a set $\mathcal{N}$ and $N-1$ branches denoted by a set $\mathcal{E}$. Let $bp(i)$ denote the bus that immediately precedes bus $i$ along the radial network headed by the feeder head bus. The three-phase indices $\phi_a,\phi_b,\phi_c$ are simplified as $\phi$. The time instance is represented by $t$. Distributed assets are located at different buses including voltage dependent ZIP loads and solar PV distributed generators. We assume that the customers are either equipped with SMs or micro-PMUs, which monitor the active and reactive load power and active PV generation power with proper time resolution. For each bus $i\in\mathcal{N}$, let $p^{\rm ZIP}_{i,\phi,t}, q^{\rm ZIP}_{i,\phi,t}\in\mathbbm{R}^{3\times1}$ denote the vector of three-phase active and reactive voltage-dependent ZIP loads at time $t$. For each bus $i\in\mathcal{G}$, let $p^{g}_{i,\phi,t}, q^{g}_{i,\phi,t}\in\mathbbm{R}^{3\times1}$ denote the vector of three-phase active and reactive power outputs of the $i$-th PV inverter at time $t$; $V_{i,\phi,t}\in\mathbbm{R}^{3\times1}$ represents the vector of three-phase voltage magnitude at time $t$, $v_{i,\phi,t}:=V_{i,\phi,t}\odot V_{i,\phi,t}\in\mathbbm{R}^{3\times1}$ represents the vector of three-phase squared voltage magnitude at time $t$. For each branch $({i,j})\in\mathcal{E}$, let $z_{ij}=r_{ij}+{\bf i} x_{ij}\in\mathbbm{C}^{3\times3}$ denotes matrix of the three-phase impedence of line $ij$, where $r_{ij}$ and $x_{ij}$ are the matrices of the three-phase resistance and reactance, respectively. Let $S_{ij,\phi,t}=P_{ij,\phi,t}+{\bf i} Q_{ij,\phi,t}\in\mathbbm{C}^{3\times1}$ denotes the vector of three-phase apparent power, where $P_{ij,\phi,t}$ and $Q_{ij,\phi,t}$ are the vector of three-phase active and reactive power flow trough line $ij$ from bus $i$ to bus $j$ at time $t$. Let the line active and reactive power flows, nodal active and reactive power injections, and squared voltage magnitudes be denoted by the following column vectors: $P :=\{P_{bp(i)i,\phi,t},\forall i,t,\phi\}$, $Q=\{Q_{bp(i)i,\phi,t},\forall i,t,\phi\}$, $p=\{p_{i,\phi,t},\forall i,t,\phi\}$, $q=\{q_{i,\phi,t},\forall i,t,\phi\}$, and $v=\{v_{i,\phi,t},\forall i,t,\phi\}$. $\odot$ and $\oslash$ denote the element-wise multiplication and division. 

The classic VVO-CVR program can be formulated as a deterministic problem (1), which aims to reduce the total power consumption of the entire distribution network while maintaining a feasible voltage profile within the predefined bounds across the distribution network as follows:
\begin{subequations}
\begin{align}
\min_{P,Q,p,q,v} \hspace{2mm}& \sum_{t\in[t,t+T]}\sum_{j:0\rightarrow j}\sum_{\phi\in\{\rm a,b,c\}}{\rm Re}\{S_{0j,\phi,t}\}\label{eq_CVR_obj},\\
\nonumber\text{s.t.}\\
P_{ij,\phi,t}&=\sum_{k:j\rightarrow k} P_{jk,\phi,t}-p^{g}_{j,\phi,t}+p_{j,\phi,t}^{\rm ZIP}+\varepsilon_{ij,\phi,t}^{p}\label{eq_CVR_P},\\
Q_{ij,\phi,t}&= \sum_{k:j\rightarrow k} Q_{jk,\phi,t}-q^{g}_{j,\phi,t}+q_{j,\phi,t}^{\rm ZIP}+\varepsilon_{ij,\phi,t}^{q}\label{eq_CVR_Q},\\ 
v_{j,\phi,t}&=v_{i,\phi,t} - 2\big(\bar{r}_{ij}\odot P_{ij,\phi,t}+\bar{x}_{ij}\odot Q_{ij,\phi,t}\big) + \varepsilon_{i,\phi,t}^{v}\label{eq_CVR_V},\\
p^{\rm ZIP}_{i,\phi,t}&= p^{\rm L}_{i,\phi,t}\odot\big(k_{i,1}^p\cdot v_{i,\phi,t}+k_{i,2}^p\cdot\sqrt{v_{i,\phi,t}}+k_{i,3}^p\big)\label{eq_CVR_ZIP_P},\\
q^{ZIP}_{i,\phi,t}&= q^{\rm L}_{i,\phi,t}\odot\big(k_{i,1}^q\cdot v_{i,\phi,t}+k_{i,2}^q\cdot \sqrt{v_{i,\phi,t}}+k_{i,3}^q\big)\label{eq_CVR_ZIP_Q},\\
-Q_{i,\phi,t}^{\rm cap}&\leq q^g_{i,\phi,t}\leq Q_{i,\phi,t}^{\rm cap}, \forall i\in\mathcal{G}\label{eq_CVR_Qgen},\\
Q_{i,\phi,t}^{\rm cap}&=\sqrt{(S^{\rm cap}_{i,\phi,t})^2-(p^{\rm g}_{i,\phi,t})^2}, \forall i\in\mathcal{G}\label{eq_CVR_Qcap},\\
v^{\rm min}&\leq v_{i,\phi,t}\leq v^{\rm max}, \forall i\in\mathcal{N}\label{eq_CVR_Vlim}.
\end{align}
\end{subequations}

In objective function \eqref{eq_CVR_obj}, the three-phase active power supplied from the substation of the feeders ${\rm Re}\{S_{0j,\phi,t}\}$ is minimized over a moving finite horizon $[t,t+T]$ for energy-saving with CVR implementation. Constraints \eqref{eq_CVR_P}-\eqref{eq_CVR_V} are defined by the unbalanced three-phase version of DistFlow model \cite{CVR_4}. Constraints \eqref{eq_CVR_P} and \eqref{eq_CVR_Q} guarantee the nodal active and reactive power balance. Constraint \eqref{eq_CVR_V} calculates the voltage difference between bus $i$ and bus $j$. The detailed formulations of nonlinear terms $\varepsilon_{ij,\phi,t}^{p}$, $\varepsilon_{ij,\phi,t}^{q}$ and $\varepsilon_{i,\phi,t}^{v}$ can be found in \cite{CVR_5}. If the network is not too severely unbalanced, the voltage magnitudes between the phases are similar and the relative phase unbalance $\alpha$ is small \cite{CVR_4}, then we can use unbalanced three-phase resistance matrix $\bar{r}_{ij}$ and reactance matrix $\bar{x}_{ij}$ in constraint \eqref{eq_CVR_V}. More details about $\bar{r}_{ij}$ and $\bar{x}_{ij}$ for unbalanced three-phase distribution systems can be referred to \cite{CVR_4}. The implementation of CVR requires the modeling of voltage-dependent ZIP active and reactive loads $p^{\rm ZIP}_{i,\phi,t}$ and $q^{\rm ZIP}_{i,\phi,t}$, as shown in \eqref{eq_CVR_ZIP_P} and \eqref{eq_CVR_ZIP_Q}. $p^{\rm L}_{i,\phi,t}, q^{\rm L}_{i,\phi,t}\in\mathbbm{R}^{3\times1}$ are the vectors of three-phase active and reactive load time-series multipliers on bus $i$ at time $t$, respectively. $k_{i,1}^p$, $k_{i,2}^p$, $k_{i,3}^p$ and $k_{i,1}^q$, $k_{i,2}^q$, $k_{i,3}^q$ are constant-impedance (Z), constant-current (I) and constant-power (P) coefficients for active and reactive ZIP loads on bus $i$. Constraint \eqref{eq_CVR_Qgen} limits the reactive power output $q^g_{i,\phi,t}$ of the PV inverters by the available reactive power capacity $Q_{i,\phi,t}^{\rm cap}$. Constraint \eqref{eq_CVR_Qcap} calculates $Q_{i,\phi,t}^{\rm cap}$ with the total capacity of the PV inverter $S^{\rm cap}_{i,\phi,t}$ and the active power output of PV inverter $p^g_{i,\phi,t}$. Based on IEEE 1547-2018 Standard \cite{IEEE_1547}, the PV inverters can provide reactive power injection or absorption $q^g_{i,\phi,t}$ to achieve fast voltage regulation. In this work, we focus on proposing a CVR model by optimally controlling the injection or absorption of reactive power in the PV inverters against the uncertainties of loads and renewable powers. While the dispatches of the on-load tap changers (OLTCs) and the capacitor banks (CBs) are slow and limited by a certain number of switching operation, which cannot response the uncertainties of loads and renewable powers. To consider the impacts of those conventional voltage regulation devices on CVR performance, a hierarchical control method \cite{CVR_5} and \cite{no_oltc_cb} can be easily implemented to coordinate PV inverters, OLTC and CBs from different control stages. Note that the coordination between PV inverters, OLTC and CBs is out of the scope of this paper. In constraint \eqref{eq_CVR_Vlim}, the squared bus voltage magnitude $v_{i,\phi,t}$ is limited by $v_{\rm min}$ and $v_{\rm max}$, which are typically $[0.95^2,1.05^2]$ p.u., respectively.

The deterministic VVO-CVR problem (1) has an underlying assumption that the load and PV generation predictions are perfect, which means $p^{\rm L}_{i,\phi,t}$, $q^{\rm L}_{i,\phi,t}$ in constraints \eqref{eq_CVR_ZIP_P} and \eqref{eq_CVR_ZIP_Q}, and $p^{\rm g}_{i,\phi,t}$ in constraint \eqref{eq_CVR_Qcap} are predefined constant parameters. The more realistic setting is to take the load and PV generation prediction errors into account. To do this, we can replace the deterministic parameters $p^{\rm L}_{i,\phi,t}$, $q^{\rm L}_{i,\phi,t}, p^{\rm g}_{i,\phi,t}$ by uncertainty variables. Particularly, we introduce the regularized uncertainty variables $p^{\rm L,\xi}_{i,\phi,t}\in[0,1]$, $q^{\rm L,\xi}_{i,\phi,t}\in[0,1]$ to replace deterministic load multipliers $p^{\rm L}_{i,\phi,t}$, $q^{\rm L}_{i,\phi,t}$, as shown in \eqref{zip_apx_3} and \eqref{zip_apx_4}. We reserve the super-script $\xi$ to define the random variables, which also apply to the rest of the definition blow. Then, we introduce an auxiliary variables $\alpha_{i,\phi,t}^{\rm q}\in[-1,1]$, which represents the ratio between reactive power output $q^{\rm g}_{i,\phi,t}$ and reactive power capacity. Here, we use $Q_{i,\phi,t}^{\rm cap,\xi}$ to represent the square root term $\sqrt{(S^{\rm cap}_{i,\phi,t})^2-(p^{\rm g,\xi}_{i,\phi,t})^2}$, so that the constraint \eqref{eq_CVR_Qcap} can be reformulated as constraint \eqref{eq3_DG_Q_1}.
\begin{subequations}
\begin{align}
p^{\rm ZIP}_{i,\phi,t}&= p^{\rm L,\xi}_{i,\phi,t}\odot\big(k_{i,1}^p\cdot v_{i,\phi,t}+k_{i,2}^p\cdot\sqrt{v_{i,\phi,t}}+k_{i,3}^p\big)\label{zip_apx_3},\\
q^{\rm ZIP}_{i,\phi,t}&= q^{\rm L,\xi}_{i,\phi,t}\odot\big(k_{i,1}^q\cdot v_{i,\phi,t}+k_{i,2}^q\cdot \sqrt{v_{i,\phi,t}}+k_{i,3}^q\big)\label{zip_apx_4},\\
q^{\rm g}_{i,\phi,t}&= \alpha_{i,\phi,t}^{\rm q}Q_{i,\phi,t}^{\rm cap,\xi}, \forall i\in\mathcal{G}\label{eq3_DG_Q_1}.
\end{align}
\end{subequations}

We can define a uncertainty variable vector $\xi_{i,\phi,t} = [(p_{i,\phi,t}^{L,\xi})^\top,(q_{i,\phi,t}^{L,\xi})^\top,(p^{\rm g,\xi}_{i,\phi,t})^\top, (Q_{i,\phi,t}^{\rm cap,\xi})^\top]^\top$ to include all the uncertainty variables. To be simplified, we avoid the indices of ${i,\phi,t}$ in vector $\xi$. To consider the impacts of uncertainty $\xi$ on voltage regulation performance, we can extend the deterministic maximum/minimum voltage constraint \eqref{eq_CVR_Vlim} to two chance constraints \eqref{eq3_v_con_chance_1} and \eqref{eq3_v_con_chance_2} as follows:
\begin{subequations}
\begin{align}
\mathbb{P}\{v - v_\textrm{max} \leq 0\} &\geq 1- \epsilon\label{eq3_v_con_chance_1}, \\
\mathbb{P}\{-v + v_\textrm{min} \leq 0\} &\geq 1- \epsilon\label{eq3_v_con_chance_2}.
\end{align}
\end{subequations}
where $\epsilon$ is a pre-defined risk level of failing to satisfy bus voltage constraint against uncertainties in $\xi$. To further make the solution robust to a group of probability distributions with controllable conservativeness, we introduce DRO and an ambiguity set of uncertainty to chance-constrained CVR and compactly formulate a DRCC-CVR problem \eqref{eq_compact_DRCC_obj}-\eqref{eq_compact_DRCC_con2} as follows:
\begin{subequations}
\begin{align}
&\underset{x}\min\hspace{2mm} \underset{\xi\sim\mathbb{P}\in\mathcal{P}}\max E_\mathbb{P}\{f(x,\xi)\}\label{eq_compact_DRCC_obj},\\
\text{s.t.}\hspace{2mm}& g_1(x) \leq 0\label{eq_compact_DRCC_con1},\\
&\mathbb{P}\left\{g_2(x,\xi) \leq 0\right\}\geq 1-\epsilon\label{eq_compact_DRCC_con2},
\end{align}
\end{subequations}
where $x$ represents the decision variable vector (i.e. reactive power dispatch of PV inverter) and $\xi\sim\mathbb{P}\in\mathcal{P}$ in objective \eqref{eq_compact_DRCC_obj} means that the uncertainty variable vector $\xi$ following the distribution $\mathbb{P}$ within an ambiguity set of distributions $\mathcal{P}$. Constraints \eqref{eq_CVR_P}-\eqref{eq_CVR_Qcap} can be represented by the compact constraint \eqref{eq_compact_DRCC_con1} and the chance constraints \eqref{eq3_v_con_chance_1} and \eqref{eq3_v_con_chance_2} can be represented by the compact constraint \eqref{eq_compact_DRCC_con2}.   

\subsection{Our Proposed Method}
The challenges of solving this DRCC-CVR problem \eqref{eq_compact_DRCC_obj}-\eqref{eq_compact_DRCC_con2} can be summarized as follows: (i) the nonlinear power flow model and voltage-dependent ZIP load model, and the chance constraints with the random variables make the DRCC-CVR problem \eqref{eq_compact_DRCC_obj}-\eqref{eq_compact_DRCC_con2} intractable to be solved; and (ii) even though we can reformulate the DRCC problem in a tractable way, the limited access to the high-resolution load and PV data will lead to an ill-posed DRCC-CVR problem and hinder the performance of CVR implementation. To address those challenges, we propose a solution to address these challenges, as shown in Fig. \ref{framework}, which includes the following two-fold:
\begin{figure}
	\vspace{-0pt} 
	\vspace{-0pt}
	\centering
	\includegraphics[width=0.8\linewidth]{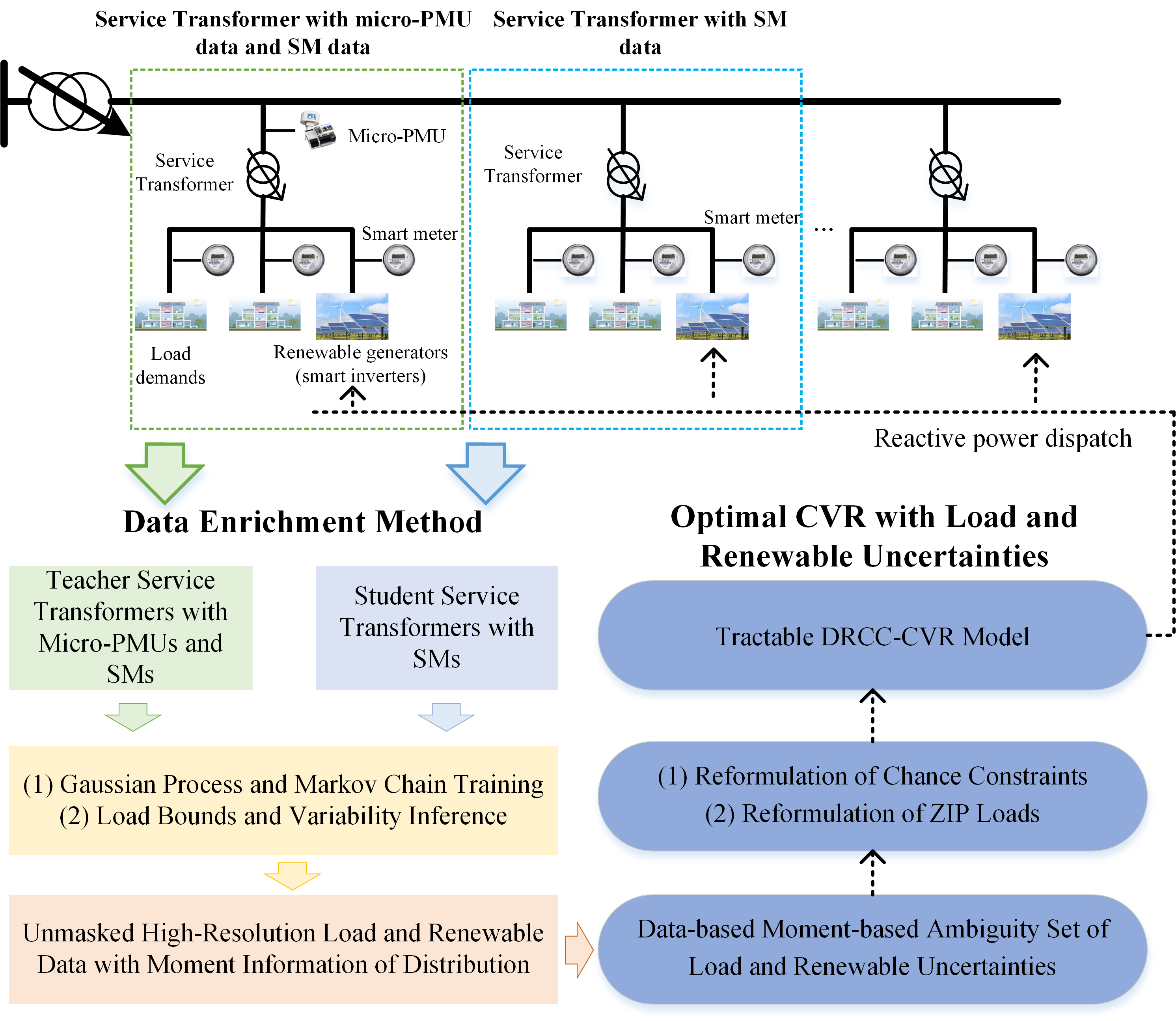}
	\vspace{-0pt} 
	\caption{Overall framework of the tractable DRCC-CVR with data enrichment method and enriched data-based moment-based ambiguity set.}
	\centering
	\label{framework}
    \vspace{-0pt} 
\end{figure} 

\textbf{1)} \textit{Tractable DRCC-CVR model:}
To reformulate a tractable DRCC-CVR model, we leverage the linearized Distflow model \cite{CVR_4} and linearized voltage-dependent ZIP load model with Binominal Approximation method \cite{ZIP_Apx_2}. Then, we reformulate the chance constraints of voltage by a tractable DRCC model with a moment-based ambiguity set. Compared to other types of DRO, the DRCC with a moment-based ambiguity set has higher computational efficiency for tractable reformulation. We will show the linearized version of the power flow model and voltage-dependent ZIP load model, as well as the tractable reformulation of the DRCC-CVR problem in Section \ref{sec:VVC}.          

\textbf{2)} \textit{Data enrichment method and moment-based Ambiguity set:}
As shown in Fig. \ref{framework}, there are a large number of service transformers that can collect low-resolution load and PV generation data by SMs, while only a few service transformers are installed with micro-PMUs with access to high-resolution load and PV generation data. To capture the uncertainty of load and PV generation, we use SMs and micro-PMUs to collect load and PV generation data. Then we enrich the load and PV generation data and extract the corresponding moment information of probability distributions from the enriched load and PV data. Finally, we can construct the ambiguity set with the first two moment information i.e., mean and variance, and implement the ambiguity set in our proposed DRCC-CVR model. The purpose of introducing the data enrichment method in a data-based ambiguity set is to avoid potential over- or under-conservativeness. We will show the data-enrichment method and the construction of a moment-based ambiguity set for DRCC-CVR in Section \ref{sec:amb_set}.     

\section{Tractable Reformulation of DRCC-CVR Model}\label{sec:VVC}
This section presents the linearized version of the power flow model and voltage-dependent ZIP load model, and reformulates a tractable version of the DDCC-CVR problem. 

\subsection{Linearized Reformulation of Power Flow and Voltage-Dependent ZIP Loads}
In power flow constraints \eqref{eq_CVR_P}-\eqref{eq_CVR_V}, the nonlinear terms $\varepsilon_{ij,\phi,t}^{p}$, $\varepsilon_{ij,\phi,t}^{p}$ and $\varepsilon_{i,\phi,t}^{v}$ make the optimization problems non-convex and NP hard. In practice, those nonlinear terms are much smaller than the linear terms in power flow constraints \eqref{eq_CVR_P}-\eqref{eq_CVR_V}. Therefore, the constraints \eqref{eq_CVR_P}-\eqref{eq_CVR_V} can be reformulated as constraints \eqref{eq_lin_Distflow_1}-\eqref{eq_lin_Distflow_3} with linearized Distflow model by neglecting those nonlinear terms. 
\begin{subequations}
\begin{align}
P_{ij,\phi,t}&=\sum_{k:j\rightarrow k} P_{jk,\phi,t}-p^{g}_{j,\phi,t}+p_{j,\phi,t}^{\rm ZIP}\label{eq_lin_Distflow_1},\\
Q_{ij,\phi,t}&= \sum_{k:j\rightarrow k} Q_{jk,\phi,t}-q^{g}_{j,\phi,t}+q_{j,\phi,t}^{\rm ZIP}\label{eq_lin_Distflow_2},\\ 
v_{j,\phi,t}&=v_{i,\phi,t} - 2\big(\bar{r}_{ij}\odot P_{ij,\phi,t}+\bar{x}_{ij}\odot Q_{ij,\phi,t}\big)\label{eq_lin_Distflow_3}.
\end{align}
\end{subequations}

This linear form of DistFlow has been verified in many previous studies, such as \cite{CVR_4}. The nonlinear term $\sqrt{v_{i,\phi,t}}$ of ZIP loads also introduces non-convexity to the problem. Because the voltage magnitudes of all buses in a distribution network stay close to 1 p.u. under normal operating conditions \cite{CVR_4} and \cite{CVR_5}, the active and reactive ZIP loads can be linearized by Binominal Approximation Method \cite{ZIP_Apx_2}. Therefore, the squared deviation of voltage $(\Delta V)^2$ is very small, so it can be neglected. Then we have the following approximations \eqref{v_apx} of squared voltage magnitude, as follows:
\begin{align}\label{v_apx}
\nonumber v&=V\odot V=(1+\Delta V)\odot(1+\Delta V)\approx 1+2\Delta V,\\
\nonumber v&=1+\Delta v\approx1+2\Delta V,\\
\Delta v&\approx2\Delta V,
\end{align}
where $v$ and $\Delta v$ are the vectors of squared voltage magnitude and the derivation from the nominal value, respectively; $V$ and $\Delta V$ are the vectors of voltage magnitude and the derivation from the nominal value, respectively. By introducing equation \eqref{v_apx} and $\sqrt{v}=V=(1+\Delta V)$ to equations \eqref{zip_apx_3} and \eqref{zip_apx_4}, we have the linear approximation of voltage-dependent active and reactive ZIP loads as follows:
\begin{subequations}
\begin{align}
p^{\rm ZIP}_{i,\phi,t}&\approx p^{\rm L,\xi}_{i,\phi,t}\odot\big((k_{i,1}^p+\frac{k_{i,2}^p}{2})v_{i,\phi,t}+(k_{i,3}^p+\frac{k_{i,2}^p}{2})\big)\label{zip_apx_1},\\
q^{\rm ZIP}_{i,\phi,t}&\approx q^{\rm L,\xi}_{i,\phi,t}\odot\big((k_{i,1}^q+\frac{k_{i,2}^q}{2})v_{i,\phi,t}+(k_{i,3}^q+\frac{k_{i,2}^q}{2})\big)\label{zip_apx_2}.
\end{align}
\end{subequations}

\subsection{Tractable Reformulation of DRCC-CVR with Load and Renewable Uncertainties}
To achieve the tractable reformulation of DRCC-CVR, the power flow constraints \eqref{eq_CVR_P}-\eqref{eq_CVR_V} can be compactly formulated as follows:
\begin{subequations}
\begin{align}
-AP&=p\label{eq_com_powerflow_1},\\
-AQ&=q\label{eq_com_powerflow_2},\\
-A_0v_0-A^\top v&=-2D_rP-2D_xQ\label{eq_com_powerflow_3},
\end{align}
\end{subequations}
where $A_0$ and $A$ are the incidence matrices of unbalanced radial distribution network, $A_0$ represents the connection structure between substation (the feeder head bus) and each of the line segments in $\mathcal{E}$, $A$ represents the connection structure between the remaining buses and each of the line segment in $\mathcal{E}$. $v_0$ is vector of square nominal voltage magnitudes. $D_r=blkdiag[R_{bp(1)1},...,R_{bp(N)N}]$ and $D_x=blkdiag[X_{bp(1)1},...,X_{bp(N)N}]$ are block diagonal matrices of line segment resistance and reactance, respectively. In equations \eqref{eq_com_powerflow_1} and \eqref{eq_com_powerflow_2}, the nodal active and reactive power injections can be calculated based on ZIP loads and PV generations. Based on the compact power flow formulations \eqref{eq_com_powerflow_1}-\eqref{eq_com_powerflow_3}, we have the compact formulation \eqref{eq_dro_1} to represent the relationship between bus voltage $v$ and bus power injections $p$ and $q$, as follows:
\begin{equation}\label{eq_dro_1}
v=Rp+Xq+\tilde{v},    
\end{equation}
with
\begin{align*}
R&=2[A^\top]^{-1}D_rA^{-1},\\
X&=2[A^\top]^{-1}D_rA^{-1},\\
\tilde{v}&=-[A^\top]^{-1}A_0v_0.    
\end{align*}

By introducing PV inverter reactive power output equation \eqref{eq3_DG_Q_1}, linearized ZIP load equations \eqref{zip_apx_1} and \eqref{zip_apx_2} as bus power injections $p$ and $q$ into the compact formulation \eqref{eq_dro_1}, we have an equation as shown in equation \eqref{v_long_1}, which represents the relationship between vector of squared voltage $v$ and all the uncertainty variables in vector $\xi$.  
\begin{figure*}[!htbp]
\hrulefill
\begin{align}\label{v_long_1}
       v = R\left(p^{L,\xi}\big((k_1^p+\frac{k_3^p}{2})v+(k_3^p+\frac{k_2^p}{2})\big)-p^{\rm g,\xi}\right)+X\left(q^{L,\xi}\big((k_1^q+\frac{k_3^q}{2})v+(k_3^q+\frac{k_2^q}{2})\big)-\alpha^{\rm q}Q^{\rm cap,\xi}\right)+\tilde{v},
\end{align}
\hrulefill
\end{figure*}
If we select an appropriate value as a power base and use per unit to represent $p^{L,\xi}$ and $q^{L,\xi}$, because the load level of distribution systems is usually not heavy at kW level, then we have $p^{L,\xi}$ and $q^{L,\xi}$ are much smaller than identity matrix $I$, the network resistance and reactance matrices $R,X$ are also small in per unit, thus the equation \eqref{val_equ} is valid. 
\begin{equation}\label{val_equ}
I - p^{L,\xi}\left(Rk_1^p + R \frac{k_2^p}{2}\right) - q^{L,\xi}\left(Xk_1^q + X \frac{k_2^q}{2}\right) > 0.
\end{equation}

After we introduce the equation \eqref{v_long_1} into the deterministic constraint (1g) on bus voltages, and because equation \eqref{val_equ} is valid, we can obtain equation \eqref{v_long_2}, which represents the impacts of load and PV generation uncertainties on bus voltage constraints (1i) with $v_{min}$ and $v_{max}$. For simplicity, we also avoid the indices of $i,\phi,t$ in equations \eqref{v_long_1} and \eqref{v_long_2}.
\begin{figure*}[!htbp]
\hrulefill
\begin{align}\label{v_long_2}
    v_{\textrm{min}} \leq\frac{p^{L,\xi}\left(Rk_3^p + R \frac{k_2^p}{2}\right)  + q^{L,\xi}\left(Xk_3^q + X \frac{k_2^q}{2}\right) + 
    Rp^{\rm g,\xi} + X \alpha^{\rm q}Q^{\rm cap,\xi} + \tilde{v}
    }{I - p^{L,\xi}\left(Rk_1^p + R \frac{k_2^p}{2}\right) - q^{L,\xi}\left(Xk_1^q + X \frac{k_2^q}{2}\right)} \leq v_{\textrm{max}},
\end{align}
\hrulefill
\end{figure*}
The compact formulation \eqref{eq_compact_DRCC_con2} can be reformulated in a linear form $a(x)^\top\xi+b(x) \leq 0$. We can obtain the formulations of $a(x)$ and $b(x)$ for chance constraints \eqref{eq3_v_con_chance_1} and \eqref{eq3_v_con_chance_2} by introducing the equation \eqref{v_long_2} into chance constraints \eqref{eq3_v_con_chance_1} and \eqref{eq3_v_con_chance_2}. Therefore, the $a(x)$ and $b(x)$ of chance constraint \eqref{eq3_v_con_chance_1} can be formulated as \eqref{a_1} and \eqref{b_1}, respectively.  
\begin{subequations}
\begin{align}
    a(x)&= \begin{bmatrix}
    \textrm{diag}(v_{\textrm{max}})\left(Rk_1^p + R \frac{k_2^p}{2}\right)+\left(Rk_3^p + R \frac{k_2^p}{2}\right)\\
    \textrm{diag}(v_{\textrm{max}})\left(Xk_1^q + X \frac{k_2^q}{2}\right)+\left(Xk_3^q + X \frac{k_2^q}{2}\right)\\
    -R \\
    -Xa_j
    \end{bmatrix}\label{a_1},\\
    b(x)&= \tilde{v} - v_{\textrm{max}}\label{b_1}.
\end{align}
\end{subequations}

Similarly, the $a(x)$ and $b(x)$ for another chance constraint \eqref{eq3_v_con_chance_2} can be formulated as \eqref{a_2} and \eqref{b_2}, respectively.
\begin{subequations}
\begin{align}
    a(x)&= \begin{bmatrix}
    -\textrm{diag}(v_{\textrm{min}})\left(Rk_1^p + R \frac{k_2^p}{2}\right)-\left(Rk_3^p + R \frac{k_2^p}{2}\right)\\
    -\textrm{diag}(v_{\textrm{min}})\left(Xk_1^q + X \frac{k_2^q}{2}\right)-\left(Xk_3^q + X \frac{k_2^q}{2}\right)\\
    R \\
    Xa_j
    \end{bmatrix}\label{a_2},\\
    b(x)&= -\tilde{v} + v_{\textrm{min}}\label{b_2}.
\end{align}
\end{subequations}

A moment-based ambiguity set of load and PV generation uncertainties can be constructed in \eqref{eq4_amb_1}, as follows:
\begin{equation}\label{eq4_amb_1}
\mathcal{D}_\xi=\left\{\xi\sim\mathbb{P}\in\mathcal{P}:\mathbb{E}_{\mathbb{P}_\xi}[\xi]=\mu,\mathbb{E}_{\mathbb{P}_\xi}[\xi\xi^T]=\Sigma\right\}
\end{equation}
where $\mu$ and $\Sigma$ represent the mean and covariance of the uncertain variables of load and PV generation. Finally, based on equations \eqref{a_1}-\eqref{b_2} for $a(x)$ and $b(x)$, we can obtain a second-order conic reformulation \eqref{eq_dro_3} for the DRCC \eqref{eq3_v_con_chance_1} and \eqref{eq3_v_con_chance_2} with moment information, mean $\mu$ and covariance $\Sigma$ of uncertainty variable vector $\xi$, as follows: \cite{delage2010distributionally} and \cite{calafiore2006distributionally}:
\begin{equation}\label{eq_dro_3}
a(x)^\top\mu+b(x)+\sqrt{\frac{1-\epsilon}{\epsilon}}||\Sigma^{\frac{1}{2}}a(x)||_2\leq 0.
\end{equation}

Even though the mean $\mu$ and covariance $\Sigma$ of load and PV generation in \eqref{eq_dro_3} can be extracted from recorded data of SMs and micro-PMUs, the reality is that we only have limited access to high-resolution load and PV generation data. The limited data will lead to the potential ill-posed of DRCC \eqref{eq_dro_3}, and further hinder the performance of DRCC-CVR. Therefore, we introduce a data enrichment method for high-resolution data recovery and ambiguity set construction in Section \ref{sec:amb_set}.

\section{Data Enrichment Method And Moment-Based Ambiguity Set}\label{sec:amb_set}
This section presents the data enrichment method to recover high-resolution data of load and PV generation for those service transformers with only SMs. Then, the ambiguity set with moment information of probability distributions is constructed based on the enriched load and PV generation data. 

\subsection{Data Enrichment Method with Micro-PMU and SM Data}
As shown in Fig. \ref{framework}, the majority of service transformers are installed with SMs to record low-resolution load and PV generation data, and only a few service transformers are also installed with micro-PMU to record high-resolution data. To enrich the data of load and PV generation, we consider the service transformers with micro-PMU data as a teacher repository and the service transformers with only SM data as a student repository. The teacher service transformers train two models capturing the statistical relationship between high-resolution data and low-resolution data. The trained models are utilized to perform data enrichment for those service transformers with only SMs. There are four main steps: 

\textbf{Step. I} \textit{Train the load/PV generation data maximum and minimum boundary inference models:} The first step is to train probabilistic models using known high-resolution load/PV generation data from teacher service transformers with micro-PMUs. The Gaussian Process Regression (GPR) technique \cite{data_enrich} is used to capture the relationship between the maximum/minimum bounds and the average values for load/PV generation. More specifically, two GPR models $GPR^*_{s,1}$ and $GPR^*_{s,2}$ are trained for the $s$-th teacher service transformer: 
\begin{subequations}\label{GPR}
\begin{align}
GPR^*_{s,1}&: P_a(t) \to \overline{P}(t)\label{GPR_1},\\
GPR^*_{s,2}&: P_a(t) \to \underline{P}(t)\label{GPR_2},
\end{align}
\end{subequations}
where $P_a(t)$ denotes the average load/PV generation over the $t$-th hour, $\overline{P}(t)$ and $\underline{P}(t)$ denote the upper and lower bounds of instantaneous load/PV generation within the $t$-th hour, respectively. 

\textbf{Step. II} \textit{Train the load/PV generation variability inference models:} The second step is to model the probabilistic transition of instantaneous load/PV generation within their maximum and minimum bounds using the second-order Markov Chain (MC) model. Specifically, one MC model $MC^*_{s}$ is trained for each service transformer:
\begin{equation}\label{MC_1}
MC^*_{s}: \{P_t(m-2), P_t(m-1)\} \to P_r(P_t(m)),
\end{equation}
where $P_t(m-2)$, $P_t(m-1)$, and $P_t(m)$ denote the $(m-2)$-th, $(m-1)$-th, and $m$-th high-resolution load/PV generation samples within the $t$-th hour. Formulation \eqref{MC_1} outputs the probability of $P_t(m)$ based on $P_t(m-2)$ and $P_t(m-1)$. Note that Step. I and II build the load/PV generation data boundary inference models and the load/PV generation data variability inference models with a small number of high-resolution micro-PMU data of teacher service transformers. Steps. III and IV will extend the trained load/PV generation probabilistic models to service transformers with only SMs, so it recovers the high-resolution load/PV generation data masked by the low-resolution load/PV generation data. 

\textbf{Step. III} \textit{Determine the learning weights of student service transformers with respect to teacher service transformers:} The third step evaluates the low-resolution data similarity between the teacher with micro-PMUs and student service transformers with SMs by determining the learning weights as shown in \eqref{eq_data_8} and in \eqref{eq_data_9}. The weights $W_s$ and $W'_s$ can represent the confidence of a student service transformer to learn from multiple teacher service transformers for enriching the low-resolution load/PV generation data to high-resolution load/PV generation data.
\begin{subequations}
\begin{align}
W_s&=\frac{W'_s}{\sum^{N_t}_{s=1}W'_s}\label{eq_data_8},\\
W'_s&=\frac{1}{N_c N^s_c}\sum^{N_c}_{i=1}\sum^{N^s_c}_{j=1}||P_i-P_j^s||, s=\{1,...,N_t\}\label{eq_data_9},
\end{align}
\end{subequations}
where $N_c$ and $N_c^s$ denote the number of customers served by a student service transformer and the the $s$-th teacher service transformer, respectively. We can obtain $N_c$ daily load/PV generation patterns for that service transformer, $\{{P}_1,\cdots,{P}_{N_c}\}$. Similarly, the load/PV generation patterns for the $s$-th teacher service transformer are denoted by $\{{P}_1^s,\cdots,{P}_{N_c^s}^s\}, s=1,\cdots,N_t$. The weights are used to linearly combining the estimated bounds in \eqref{GPR} and the probabilistic transition matrices in \eqref{MC_1}.

\textbf{Step. IV} \textit{Extend the trained load/PV generation data probabilistic model:} The fourth step extends the trained probabilistic models of teacher service transformers in Steps. I and II to student service transformers that only have SMs for enriching low-resolution load/PV generation data. Specifically, the $m$-th high-resolution load/PV generation sample is randomly generated based on the following Bernoulli distribution:
\begin{equation}\label{generate_data}
{P}_t(m)\sim Be(Pr({P}_t)),
\end{equation}
where $Be$ denotes the Bernoulli distribution \cite{data_enrich}. Note that for each hourly load/PV generation sample, \eqref{generate_data} can give us $N$' high-resolution load/PV generation samples , i.e., $P_t(m), m=1,\cdots,N'$. Then, the enriched high-resolution data samples are employed to optimize the mean and standard deviation of Gaussian distribution using maximum likelihood estimation:
\begin{equation}\label{MLE_for_mu_sigma}
(\mu^*, \Sigma^*) = \underset{\mu, \Sigma}{arg min} \;
\, \prod_{m=1}^{N'} f(\mu, \Sigma; P_t(m)),
\end{equation}
where, $f(\cdot)$ denotes the probability density function of Gaussian distribution. Therefore, the mean and covariance of load and PV generation uncertainties $\mu$ and $\Sigma$ can be execrated from enriched data in \eqref{MLE_for_mu_sigma} for the construction of the moment-based ambiguity set \eqref{eq4_amb_1}. i.e., substitute $\mu^*$ and $\Sigma^*$ in \eqref{MLE_for_mu_sigma} into \eqref{eq_dro_3}.   

\subsection{Enriched Data-Based Ambiguity Set and DRCC-CVR Method}
The enriched high-resolution data from \eqref{generate_data} can recover instantaneous uncertainties of load and PV generation, which can be further extracted for the two moment information and construct moment-based ambiguity set in \eqref{eq4_amb_1}. By considering the data-based moment-based ambiguity set and the reformulation of tractable DRO and chance constraints, the DRCC-CVR problem can be compactly reformulated as follows:
\begin{subequations}
\begin{align}
&\underset{\xi\in\mathcal{D}_\xi}{\text{Objective}}\hspace{2mm} (1),\label{eq_compact_CVR_obj}\\
\text{s.t.}\hspace{2mm}& \text{PV generations}\left\{(2c)\right\},\label{eq_compact_CVR_1}\\
&\text{OPF constraints}\left\{\text(5)\right\},\label{eq_compact_CVR_2}\\
&\text{Linearized ZIP loads}\left\{(7)\right\},\label{eq_compact_CVR_3}\\
&\text{Tractable reformulation of DRCC}\left\{\text (13)-(14),(16)\right\}.\label{eq_compact_CVR_4}
\end{align}
\end{subequations}

Note that problem \eqref{eq_compact_CVR_obj}-\eqref{eq_compact_CVR_4} is a tractable linear programming problem, where the mean values of load and PV are used in compact constraints \eqref{eq_compact_CVR_1}-\eqref{eq_compact_CVR_3} and the mean and covariance values of load and PV are used in the compact constraint \eqref{eq_compact_CVR_4} of DRCC reformulation. 
To summarize the above steps, the detailed procedure of the proposed DRCC-CVR with an enriched-based ambiguity set of the uncertainties of load and PV generation is shown in Algorithm \ref{alg:Data_enrich}.
\begin{algorithm}[t]
\caption{DRCC-CVR Model with Enriched-based Ambiguity Set of Uncertainty of Load and PV Generation}\label{alg:Data_enrich}
\begin{algorithmic}[1]
  \State {\bf Input}: High-resolution data from micro-PMUs and low-resolution data from SMs
  \State \hspace{0mm}{\bf Initialization}: Choose hyper-parameters in DRCC-CVR 
  \State {\bf For}: $i=1,2,...,N$.
  \State Train load/PV generation upper and lower boundary inference models from teacher service transformers with micro-PMUs by \eqref{GPR_1} and \eqref{GPR_2}.
  \State Train load/PV generation variability inference model from teacher service transformers with micro-PMUs by \eqref{MC_1}.
  \State Determine the learning weights of student service transformers with respect to teacher service transformers in \eqref{eq_data_8} and \eqref{eq_data_9}.
  \State Extend the trained load/PV generation data to student service transformers with SMs in \eqref{generate_data}.
  \State {\bf End for}.
  \State Extract the first two moment information of load and PV generation uncertainties from enriched data in \eqref{MLE_for_mu_sigma} 
  \State Construct an ambiguity set with the first two moment information in \eqref{eq4_amb_1}.
     \State Solve the DRCC-CVR problem (22) with objective (1) and constraints (2c),(5),(7), (13)-(14), and (16).
    \State {\bf Output}: Reactive power dispatches of PV inverters
\end{algorithmic}
\end{algorithm}

\section{Case Studies}\label{sec:Results}
This section presents the simulation results, including the enriched data of load and PV generation for ambiguity set, comparison results of the benchmark methods and the proposed DRCC-CVR method, and the impacts of hyper-parameter on the proposed DRCC-CVR method.  

\subsection{Simulation Setup}
\begin{figure}
	\vspace{-0pt} 
	\vspace{-0pt}
	\centering
	\includegraphics[width=0.8\linewidth]{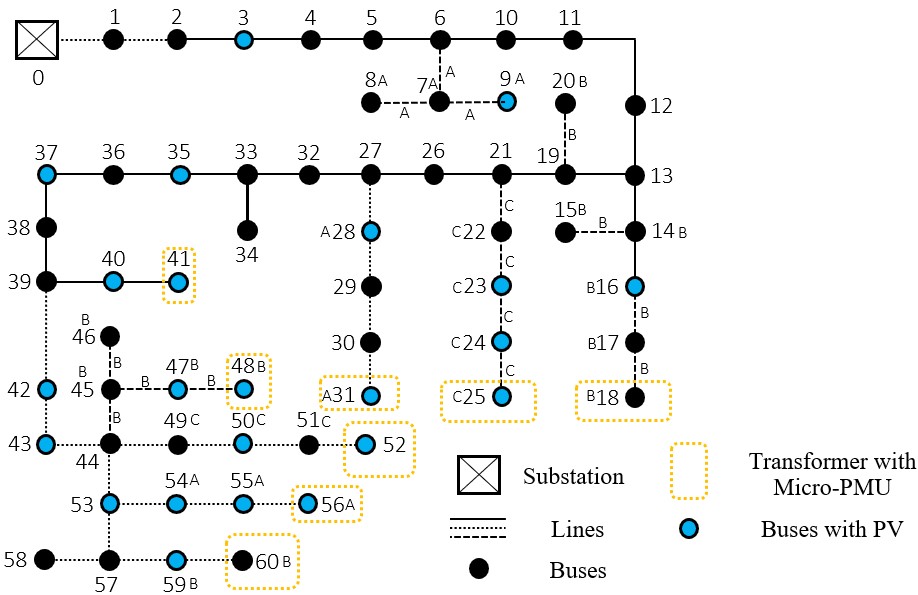}
	\vspace{-0pt} 
	\caption{A real distribution feeder in Midwest U.S. \cite{Realsystem}}
	\centering
	\label{test}
    \vspace{-0pt} 
\end{figure}

A real-world distribution feeder \cite{Realsystem} in Fig. \ref{test} is used to test our proposed DRCC-CVR method, which is located in Midwest U.S. and shared by our utility partner. The reason for choosing this real-world distribution feeder as our test system is that the service transformers of customers in this feeder are either equipped with SMs or micro-PMUs, which can record low-resolution (1-hour) data and high-resolution (1-second) data to construct an ambiguity set of load and PV generation uncertainties for our DRCC-CVR method. More information on this real-world distribution feeder and the data from SMs and micro-PMUs can be found in \cite{Realsystem}. In Fig. \ref{test}, the yellow bot-ted boxes represent the buses' service transformers installed with micro-PMUs, and the rest buses' service transformers are installed with SMs; the blue dots represent the buses installed with single-phase or three-phase PV generators; the solid, dashed and dotted lines represent three-phase overhead lines, three-phase underground cables and single-phase overhead lines, respectively. In this real-world distribution feeder, the total capacity of PVs can serve around 20\% to 30\% load. Adopted from our industrial partner \cite{CVR_5}, we use the following coefficients $[k_{1}^p,k_{2}^p,k_{3}^p]=[0.96, -1.17, 1.21]$ and $[k_{1}^q,k_{2}^q,k_{3}^q]=[6.28, -10.16, 4.88]$ for active and reactive ZIP loads in our test cases. The base voltage and base power values are 13.8 kV and 100 kVA. The prescribed risk level parameter $\epsilon$ is set to 0.05 for quantifying the 5\% violation probability of chance constraints in our proposed DRCC-CVR model.
We demonstrate the advantages and effectiveness of the enriched data-based ambiguity set and the proposed DRCC-CVR method through numerical comparisons of several benchmark methods. The following simulations are built-in MATLAB R2019b, which integrate YALMIP Toolbox with IBM ILOG CPLEX 12.9 solver for optimization. All case studies are simulated on a PC with Intel Core i7-4790 3.6 GHz CPU and 16 GB RAM.  

\subsection{Original and Enriched Data of Load and PV Generation}\label{sec:data}
\begin{figure}
	\vspace{-0pt} 
	\vspace{-0pt}
	\centering
	\includegraphics[width=0.7\linewidth]{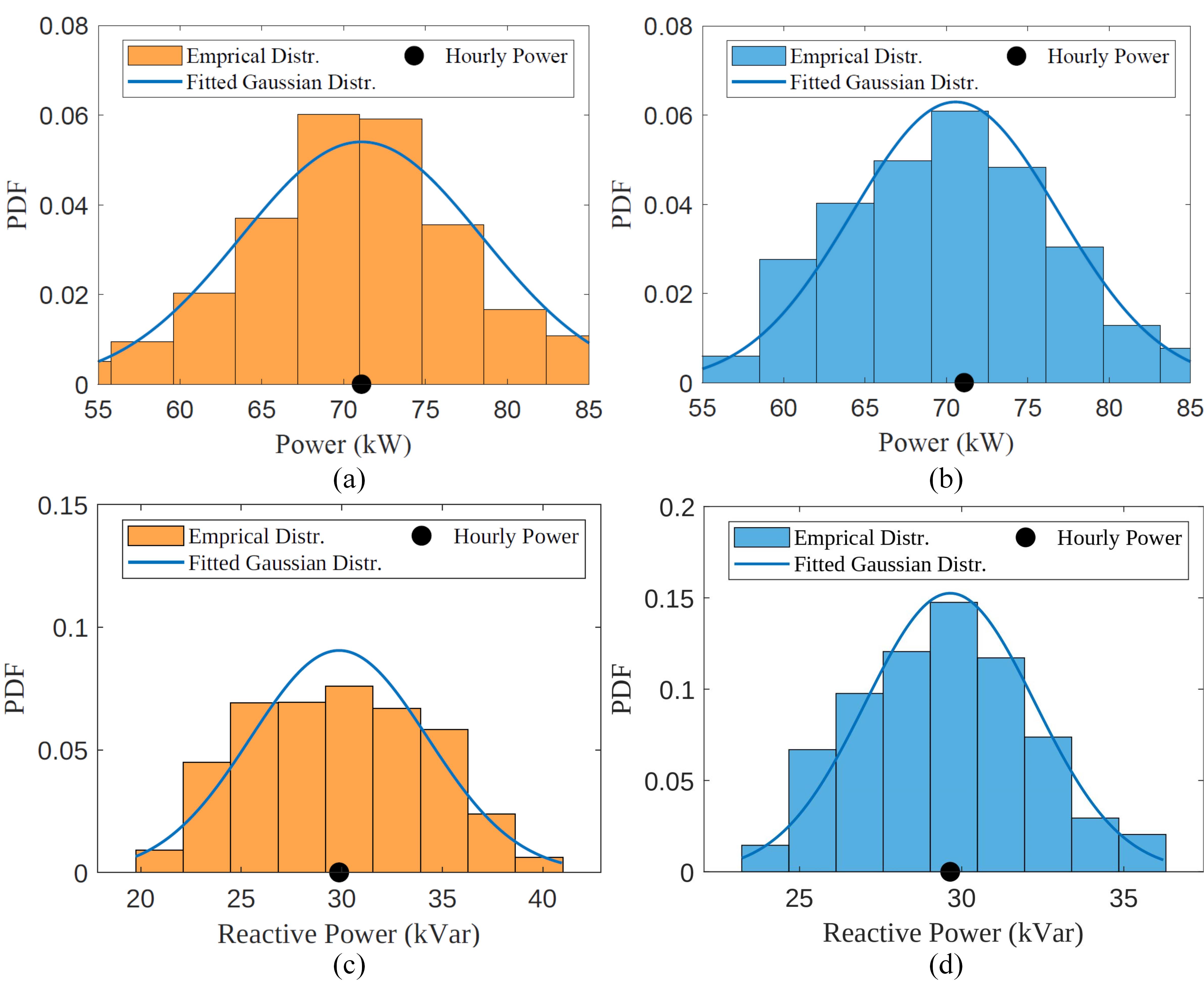}
	\vspace{-0pt} 
	\caption{Empirical distribution and fitted Gaussian distribution: (a) active load (original data); (b) active load (enriched data); (c) reactive load (original data); (d) reactive load (enriched data).}
	\centering
	\label{PQ_distribution}
    \vspace{-0pt} 
\end{figure}

\begin{figure}
	\vspace{-0pt} 
	\vspace{-0pt}
	\centering
	\includegraphics[width=0.7\linewidth]{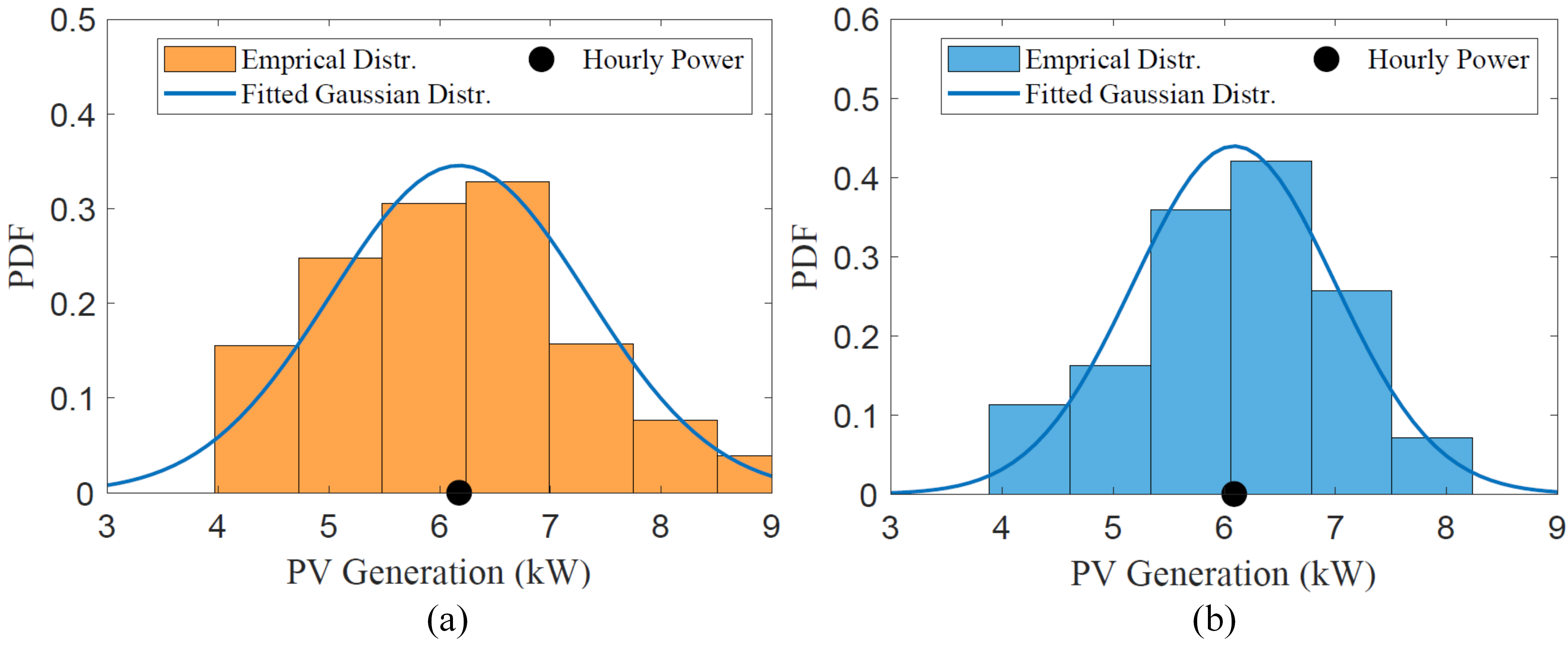}
	\vspace{-0pt} 
	\caption{Empirical distribution and fitted Gaussian distribution: (a) PV generation (original data); (b) PV generation (enriched data).}
	\centering
	\label{PV_distribution}
    \vspace{-0pt} 
\end{figure}

As shown in Fig. \ref{test}, there are 8 service transformers installed with micro-PMUs and the rest 34 service transformers installed with SMs, which can record high-resolution (1-second) and low-resolution (1-hour) load and PV generation data, respectively. In this sub-section, we use two ways to obtain the mean and covariance of the uncertainty variables of load and PV generation for ambiguity sets: (i) we use a statistical method to obtain mean and variance information of the original load and PV generation data (few high-resolution data from 8 micro-PMUs and a lot of low-resolution data from 34 SMs). (ii) We use the data enrichment method to enrich the original load and PV generation data, then we obtain mean and variance information of the enriched data. To verify the effeteness of the data enrichment method, we show the empirical distributions and their according fitted Gaussian distributions of active load, reactive load, and active power output of PV generation in Fig. \ref{PQ_distribution} and Fig. \ref{PV_distribution}, respectively. Note that the distributions in the right column and left column of Fig. \ref{PQ_distribution} and Fig. \ref{PV_distribution} are obtained from original data and enriched data, respectively. In Fig. \ref{PQ_distribution} and Fig. \ref{PV_distribution}, each dot denotes the hourly average power corresponding to the data samples, which means the hourly power is the average of observations. It can be observed that the distributions obtained from enriched data have a highly similar fitting to the distributions obtained from original data. Therefore, the first two moment information extracted from enriched data can also accurately construct the ambiguity set for the DRCC-CVR method. 

\subsection{Voltage Reduction and Power/energy-saving Through CVR Implementation}\label{sec:compare}
\begin{figure}
	\vspace{-0pt} 
	\vspace{-0pt}
	\centering
	\includegraphics[width=0.7\linewidth]{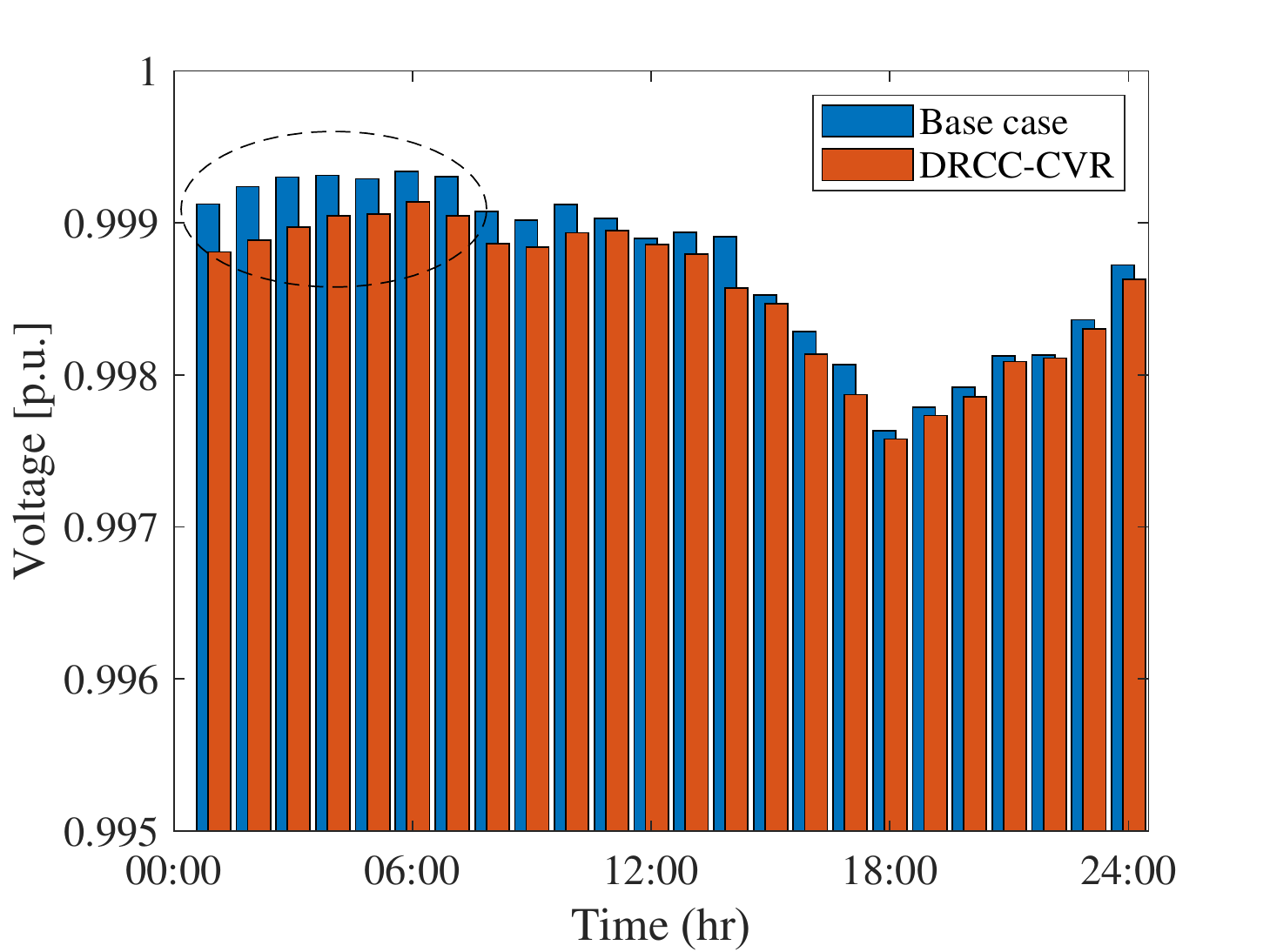}
	\vspace{-0pt} 
	\caption{Voltage profiles on selected bus 23 on phase c with different control strategies.}
	\centering
	\label{V_23b_phaseC}
    \vspace{-0pt} 
\end{figure}

To serve as a reference to investigate the performance of VVO-CVR, a base case is firstly built by setting the unity-power factor control model for all PV generators, which means there is no reactive power support from PV generators. In this sub-section, we use three ways to implement VVO-CVR: (i) The deterministic VVO-CVR (Deter-CVR) is solved by deterministic optimization, where the load and PV prediction errors and uncertainties are neglected. (ii) The robust VVO-CVR (RO-CVR) is solved by the robust optimization \cite{RO_VVC_1} and \cite{RO_VVC_2}, where the uncertainties of load and PV generation are considered with 10\% variance from the predictions. (iii) The VVO-CVR is solved by our proposed DRCC-CVR with an enriched data-based ambiguity set of load and PV generation uncertainties. The performance through CVR implementation can be evaluated from three aspects: voltage profile, active power supply from the substation, and total energy consumption. To show the time-series simulation, the VVO-CVR is performed in a daily operation of the real-world distribution feeder with different control strategies. In Fig. \ref{V_23b_phaseC}, the voltage profiles for a selected bus (bus 23 on phase c) are shown, which are generated from the base case (without control) and the proposed DRCC-CVR. In Fig. \ref{V_23b_phaseC}, the blue bar and red bar represent the voltage profiles of the base case and DRCC-CVR, respectively. It can be observed that all the nodal voltages can maintain within the predefined range [0.95,1.05] p.u., while the voltage profiles of DRCC-CVR are overall lower than the voltage profiles of the base case. Because the DRCC-CVR can optimally dispatch the reactive power from PV inverters to achieve maximum voltage reduction while still satisfying the voltage constraints. During the midnight period from 00:00 to 6:00, the voltage reduction of DRCC-CVR is obviously higher than the voltage reduction of the base case, as shown in the circled part in Fig. \ref{V_23b_phaseC}. It is because the active power outputs of PV generators from 00:00 to 6:00 are nearly zero, and according to the calculations of reactive power output and capacity in constraints \eqref{eq_CVR_Qgen} and equation \eqref{eq_CVR_Qcap}, the DRCC-CVR has more reactive power supports to maximize voltage reductions.    

\begin{figure}
	\vspace{-0pt} 
	\vspace{-0pt}
	\centering
	\includegraphics[width=0.75\linewidth]{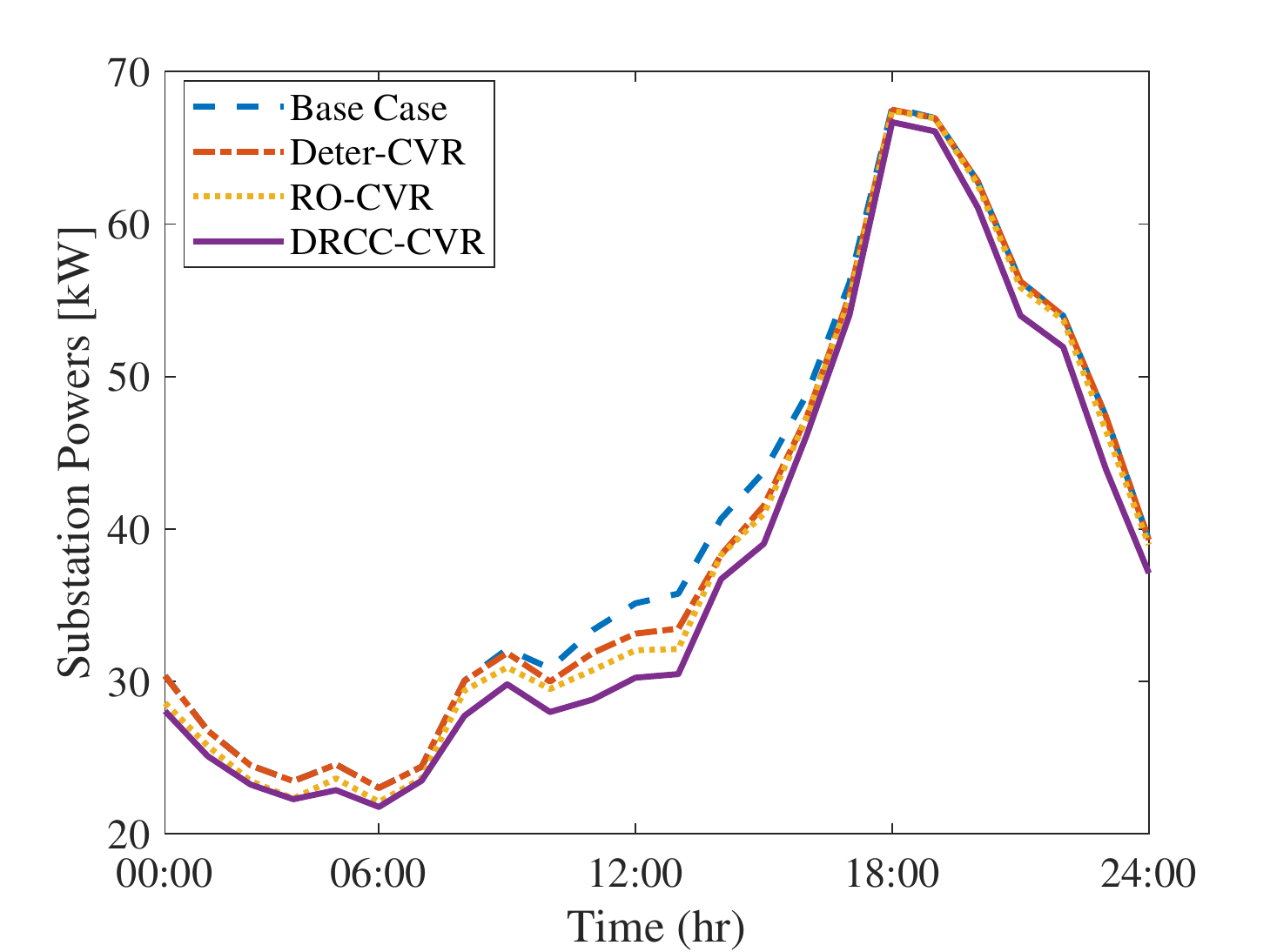}
	\vspace{-0pt} 
	\caption{Substation feed-in active power with different control strategies.}
	\centering
	\label{Psub}
    \vspace{-0pt} 
\end{figure}

The active power supplies from the substation of the base case, Deter-CVR, RO-CVR, and DDRC-CVR are shown as different curves in Fig. \ref{Psub}, which represent the overall active power consumption of the base case, and those VVO-CVR benchmarks. It can be observed that the proposed DDRC-CVR can effectively reduce the power supply from the substation compared to the base case and other methods. In comparison the power saving of RO-CVR is less than the proposed DDRC-CVR and only slightly better than Deter-CVR. On the one hand, the proposed DDRC-CVR has an ambiguity set of uncertainties to balance the trade-off between the conservativeness of decisions and operational efficiency. On the other hand, RO-CVR is too conservative and hinders the performance of CVR implementation. The numerical comparisons of total energy consumption over one day and the energy reduction percentage are presented in Table \ref{table_energy_saving_1} among base case, Deter-CVR, RO-CVR, and DDRC-CVR. The total energy consumption of base case, Deter-CVR, RO-CVR, and DRCC-CVR are 958.045 kWh, 944.048 kWh, 934.178 kWh, and 898.616 kWh, respectively. Therefore, compared to the original energy consumption in the base case, the Deter-CVR, RO-CVR, and DRCC-CVR can achieve 1.461\%, 2.491\%, and 6.203\% of energy savings, respectively. The computational times of Deter-CVR, RO-CVR, and DRCC-CVR for this test system are 12.968 seconds, 18.312 seconds, and 21.911 seconds, respectively. According to the above results in Fig. \ref{V_23b_phaseC}, Fig. \ref{Psub} and Table \ref{table_energy_saving_1}, we can summarize the differences between Deter-CVR, RO-CVR, and DRCC-CVR: (i) Among all the methods, DRCC-CVR can achieve the lowest total energy consumption and highest energy-saving. While the differences between the total energy consumption and energy-saving results of Deter-CVR and RO-CVR are small, because Deter-CVR does not consider the errors of load and PV generation predictions, and RO-CVR provides a conservative solution. Both Deter-CVR and RO-CVR cannot fully explore the benefit of CVR implementation, while the proposed DDRC-CVR can better explore the potential of CVR implementation. (ii) Because of the different ways of handling uncertainties of load and PV generation, the computation time of DRCC-CVR is slightly slower than Deter-CVR and RO-CVR. However, their differences in computational time are very small, which can be neglected for a day-ahead operational application. Thus the computational efficiency of the DRCC-CVR can be acceptable.   

\begin{table}[]
		\centering
		\renewcommand{\arraystretch}{1.3}		
		\caption{Energy Consumption and energy-saving Results With Different Control Strategies}
		\label{table_energy_saving_1}
			\vspace{-0pt}\
\begin{tabular}{lccc}
\hline\hline
                                   & Energy (kWh)      & Reduction (\%) & Computation (sec)\\\hline
Base Case                          & 958.045           &   -            &   -            \\
Deter-CVR                          & 944.048           & 1.461\%        & 12.968       \\
RO-CVR                             & 934.178           & 2.491\%        & 18.312        \\
DRCC-CVR                           & 898.616           & 6.203\%        & 21.911        \\
\hline\hline
\end{tabular}
\end{table}

\subsection{Impact of Hyper-Parameters in Performance of DRCC-CVR}\label{sec:PMU}
It is obvious that more micro-PMUs installed in service transformers can record more high-resolution data of load and PV generation. Also, more high-resolution data can be beneficial for recording the uncertainties of load and PV generation, which is helpful for the data enrichment method and construction of ambiguity set in DDRC-CVR. To show the impact of micro-PMUs on the proposed DRCC-CVR, the total energy consumption and energy-saving of DRCC-CVR with the different numbers of micro-PMUs are presented in Table \ref{table_energy_saving_2}. There are three different tests: the first test only constructs the ambiguity set based on low-resolution data from SMs without any high-resolution data from micro-PMU; the second test implements the data enrichment method with high-resolution data from 4 micro-PMUs and low-resolution data from SMs, then the ambiguity set is constructed based on the enriched data; the third test has the high-resolution data from 8 micro-PMUs as the same setting in Section V-B and Section V-C. It can be observed that a higher number of micro-PMUs can help to construct a high-quality ambiguity set for DRCC-CVR and achieve better performance of CVR implementation through DRCC-CVR.  
\begin{table}[]
    \centering
	\renewcommand{\arraystretch}{1.3}	
	\caption{Energy Consumption and energy-saving Results With Different Number of Micro-PMUs.}
	\label{table_energy_saving_2}
	\begin{tabular}{lccc}
		\hline\hline
		\begin{tabular}[l]{@{}l@{}}Number of \\ micro-PMUs \end{tabular}                     &
		\begin{tabular}[l]{@{}l@{}}Location of \\ micro-PMUs\end{tabular}                    &
		\begin{tabular}[l]{@{}l@{}}Energy \\ (kWh)\end{tabular}                          & \begin{tabular}[l]{@{}l@{}}Saving \\ (\%)\end{tabular}                           \\                  
	    \hline
		0 (only SM data)        &  - 
		                        &  933.889   
		                        &  2.521\%   \\
		\hline
	    4                       & \begin{tabular}[l]{@{}l@{}}$B_{18},B_{25},B_{41},B_{60}$\end{tabular} 
	                            &  927.817   
	                            &  3.155\%  \\
	    \hline
	    8                       & \begin{tabular}[l]{@{}l@{}}$B_{18},B_{25},B_{31},B_{41}$\\
		                                                     $B_{48},B_{52},B_{56},B_{60}$\end{tabular}   
		                        &  898.618   
	                            &  6.203\%  \\
	    \hline\hline
	\end{tabular}
\end{table}

The pre-defined risk level $\epsilon$ is selected as 0.05 in the above simulation tests. While the different values of $\epsilon$ will influence the confidence level on chance constraints in the proposed DRCC-CVR and further influence the benefits of CVR implementation. Therefore, we also test the DRCC-CVR with three different violation rates (i.e. 0.02, 0.05, 0.1), as shown in Table \ref{table_energy_saving_3}. It can be observed that a larger value of $\epsilon$ leads to lower energy consumption and a higher energy reduction, which will benefit the CVR implementation. However, this kind of benefit is achieved by increasing the violation rate and sacrificing the reliability of operational constraints. For example, the energy reduction ($6.606\% $) of $\epsilon=0.1$ is only slightly higher than the energy reduction ($6.203\% $) with $\epsilon=0.1$, but the risk of constraint violation also become higher when $\epsilon$ increases from $0.05$ to $0.1$. The results indicate that there is a trade-off between the maximization of CVR benefit and the reliability of operational constraints in the proposed DRCC-CVR.  
\begin{table}[]
		\centering
		\renewcommand{\arraystretch}{1.3}		
		\caption{Energy Consumption and energy-saving Results With Different Violation Rates}
		\label{table_energy_saving_3}
			\vspace{-0pt}\
\begin{tabular}{lcc}
\hline\hline
                                   & Energy (kWh)      & Reduction (\%) \\\hline
$\epsilon=0.02$                      & 904.980           & 4.803\%        \\
$\epsilon=0.05$                      & 898.616           & 6.203\%        \\
$\epsilon=0.1$                       & 894.754           & 6.606\%        \\
\hline\hline
\end{tabular}
\end{table}

\section{Conclusion}\label{sec:Con}
To better consider the impacts of load and PV generation uncertainties on voltage regulating performance while implementing CVR in unbalanced three-phase distribution systems, a DRCC-CVR model with an enriched data-based and moment-based ambiguity set is proposed to optimally dispatch the reactive power of PV inverters. The original and intractable DRCC-CVR is approximated by linearized ZIP load models and reformulated in a tractable way with the first two moment information of load and PV generation uncertainties. We further implement a data enrichment method with low-resolution data from SMs and high-resolution data from micro-PMUs to recover instantaneous uncertainties of load and PV generation. An ambiguity set is constructed based on enriched data for DRCC-CVR. Simulation results on a real Midwest U.S. distribution feeder have validated the effectiveness and robustness of the proposed DRCC-CVR. According to the case studies, we have shown that: (i) The data enrichment can construct an accurate ambiguity set of load and PV generation uncertainties. (ii) Compared to other benchmark methods, the proposed DRCC-CVR shows a better trade-off between the conservativeness of decisions and operational efficiency. Thus, the proposed DRCC-CVR can achieve better CVR performance on voltage reduction and power/energy-saving. (iii) With a reasonable number of micro-PMUs, a high-quality ambiguity set can be constructed for DRCC-CVR and better performance of CVR can be achieved. (iv) To consider the benefits of CVR implementation and the reliability of operational constraints, a proper risk and confidence level needs to be tuned in DRCC-CVR.






\ifCLASSOPTIONcaptionsoff
  \newpage
\fi



\bibliographystyle{IEEEtran}
\bibliography{IEEEabrv,./bibtex/bib/IEEEexample}

\end{document}